\documentclass[aps, prb, reprint, superscriptaddress, amsmath, amssymb, showkeys]{revtex4-2}

\usepackage{graphicx}
\usepackage{dcolumn}
\usepackage{bm}
\usepackage{soul}
\usepackage{color}
\usepackage{mathtools}
\usepackage[colorlinks=true, allcolors=blue]{hyperref}

\DeclareUnicodeCharacter{2061}{}

\begin{document}

\title{Magnetic critical phenomena and low temperature re-entrant spin-glass features of Al$_2$MnFe Heusler alloy}

\author{Abhinav Kumar Khorwal}
\affiliation{Department of Physics, Central University of Rajasthan, Ajmer - 305817, Rajasthan, India}

\author{Sujoy Saha}
\affiliation{Department of Physics, Central University of Rajasthan, Ajmer - 305817, Rajasthan, India}

\author{Mukesh Verma}
\affiliation{Department of Physics, Central University of Rajasthan, Ajmer - 305817, Rajasthan, India}

\author{Lalita Saini}
\affiliation{Department of Physics, Indian Institute of Technology Gandhinagar, Palaj, Gandhinagar - 382055, Gujarat, India}

\author{Suvigya Kaushik}
\affiliation{Department of Physics, Indian Institute of Technology Gandhinagar, Palaj, Gandhinagar - 382055, Gujarat, India}

\author{Yugandhar Bitla}
\affiliation{Department of Physics, Central University of Rajasthan, Ajmer - 305817, Rajasthan, India}

%\author{Alexey V. Lukoyanov}
%\affiliation{M.N. Mikheev Institute of Metal Physics UrB RAS, 620108, Ekaterinburg, Russia}
%\affiliation{Institute of Physics and Technology, Ural Federal University, 620002, Ekaterinburg, Russia}

\author{Ajit K Patra}
\email[Corresponding author: ]{akpatra@uohyd.ac.in}
\affiliation{Department of Physics, Central University of Rajasthan, Ajmer - 305817, Rajasthan, India}
\affiliation{School of Physics, University of Hyderabad, Hyderabad - 500046, Telangana, India}

\begin{abstract}
A detailed investigation of the structural and magnetic properties, including magnetocaloric effect, re-entrant spin-glass behavior at low temperature, and critical behavior in polycrystalline Al$_2$MnFe Heusler alloy is reported. The prepared alloy crystallizes in a cubic CsCl-type crystal structure with Pm$\Bar{3}$m space group. The temperature-dependent magnetization data reveals a second-order paramagnetic to ferromagnetic phase transition ($\sim$ 122.9 K), which is further supported by the analysis of the magnetocaloric effect. The isothermal magnetization loops show a soft ferromagnetic behavior of the studied alloy and also reveal an itinerant character of the underlying exchange interactions. In order to understand the nature of magnetic interactions, the critical exponents for spontaneous magnetization, initial magnetic susceptibility, and critical MH isotherm are determined using Modified Arrott plots, Kouvel-Fisher plots, and critical isotherm analysis. The derived critical exponents $\beta$ = 0.363(2), $\gamma$ = 1.384(3), and $\delta$ = 4.81(3) confirm the critical behavior similar to that of a 3D-Heisenberg-type ferromagnet with short-range exchange interactions that are found to decay with distance as J(r) $\approx$ r$^{-4.936}$. Moreover, the detailed analysis of the AC susceptibility data suggests that the frequency-dependent shifting of the peak temperatures is well explained using standard dynamic scaling laws such as the critical slowing down model and Vogel-Fulcher law, and confirms the signature of re-entrant spin-glass features in Al$_2$MnFe Heusler alloy. Furthermore, maximum magnetic entropy change of $\sim$ 1.92 J/kg-K and relative cooling power of $\sim$ 496 J/kg at 50 kOe applied magnetic field are determined from magnetocaloric studies that are comparable to those of other Mn-Fe-Al systems. Although the values are lesser than those of pure Gd and other costly rare-earth based magnetocaloric materials, the studied alloy shows remarkable magnetocaloric properties that can be harnessed for cost-effective applications.
\end{abstract}


\maketitle

\section{\label{sec:Introduction}Introduction}

Designing novel materials with tunable and diverse functionalities is one of the core activities of physicists and materials scientists.  In this context, Heusler alloys (HAs), ternary intermetallic alloys with a general formula X$_2$YZ (X and Y are transition metals, and Z is a main group element), have gathered significant attention because innumerable alloys can be designed using almost all the elements of the periodic table \cite{ref1}. HAs adopt various crystal structures and possess a wide range of ground states such as ferromagnetism, antiferromagnetism, ferrimagnetism, half-metallic ferromagnetism, skyrmion, topological insulating nature, superconductivity, spin gapless semiconducting behavior, thermoelectric effects, topological Hall effect, spin Hall effect, etc \cite{ref2}. Due to their chemical flexibilities and exotic behavior, HAs are ideal systems for crystal engineering for various technological applications, and are interesting systems for investigating the nature of magnetic correlations. Full Heusler alloys (FHAs) crystallize either in an L2$_1$-structure (Cu$_2$MnAl-type: Fm$\Bar{3}$m space group) or an XA-structure (Hg$_2$CuTi-type: F$\Bar{4}$3m space group) \cite{ref3}. The XA structure is formed when the number of valence electrons of Y atoms is higher than that of X atoms. The unit cell of a HA consists of four interpenetrating face-centered-cubic sublattices (A, B, C and D) with Wyckoff coordinates (0, 0, 0), (1/4, 1/4, 1/4), (1/2, 1/2, 1/2)  and (3/4, 3/4, 3/4), respectively. The L2$_1$-structure has two X atoms at A and C sites, whereas Y and Z atoms occupy B and D sites, respectively. In contrast, the XA-structure has two X atoms at A and B sites, and Y and Z atoms at C and D sites, respectively \cite{ref4}. FHAs with stoichiometric formula X$_2$YZ or Y$_2$XZ are extensively investigated \cite{ref1}.

Recently, HAs with the reverse stoichiometric formula Z$_2$XY having a cubic structure are reported for Ga$_2$MnNi \cite{ref5}, Ga$_2$FeNi \cite{ref6}, Ga$_2$MnCo \cite{ref7}, and Ga$_2$MnPd, Al$_2$MnCo, Al$_2$MnPd \cite{ref8}. In spite of a reverse chemical formula,  Ga$_2$MnCo and  Ga$_2$MnPd crystallize in an L2$_1$-structure. However, Al$_2$MnCo and Al$_2$MnPd adopt a CsCl-type cubic structure. These FHAs show two magnetic transitions - a paramagnetic (PM) to ferromagnetic (FM) phase transition at higher temperature and a glassy phase at low temperature \cite{ref7, ref8}. A detailed investigation using neutron diffraction study reveals that the formation of low temperature glassy phase in Ga$_2$MnCo can be ascribed to the coexistence of long-range FM order and site disorder (10\% swapping between Mn and Co positions) induced antiferromagnetic (AFM) correlations \cite{ref7}.

Mn-Fe-Al based systems are another interesting class of HAs on account of their diverse structural and magnetic phases. Due to the presence of two magnetic atoms with chemical similarity and their abundance, they form an exciting family of HAs to explore and develop novel materials with extraordinary physical properties.

Using mean-field renormalized-group approach, Zamora et al., have investigated the magnetic ground states of Mn-Fe-Al alloys in the disordered bcc phase. Considering only the magnetic interactions between Fe-Fe (FM), Fe-Mn (AFM) and Mn-Mn (AFM) ions, the random-bond Ising model predicts FM, AFM, and low-temperature spin-glass (SG) features for Fe-rich, Mn-rich and intermediate Fe-Mn concentration in the Mn-Fe-Al alloys. These exotic magnetic phases are reported for Fe$_x$Mn$_{0.3}$Al$_{0.7-x}$, Fe$_{0.89-x}$Mn$_{0.11}$Al$_x$, Fe$_x$Mn$_{0.6-x}$Al$_{0.4}$ alloys \cite{ref9, ref10, ref11}. One of the members, Fe$_2$MnAl shows a FM ground state with an L2$_1$ type cubic structure \cite{ref12, ref13}. On the other hand, Mn-rich Mn$_2$FeAl alloys, although theoretically predicted to possess an XA-type structure with a FM phase \cite{ref14}, are found to crystallize in a $\beta$-Mn structure with an AFM ground state and exhibit SG behavior at low temperatures for experimentally synthesized polycrystalline samples \cite{ref15, ref16, ref17}. These observations are consistent with the theoretical calculations, which suggest that with the expanded unit cell volume, the $\beta$-Mn structure is energetically more favorable than the XA-type structure \cite{ref17}. Our other works on Mn-rich Mn$_{1.5}$Fe$_{1.5}$Al \cite{ref18} and Mn$_2$Fe$_{1+x}$Al$_{1-x}$ \cite{ref19} reveal that these alloys also have a $\beta$-Mn crystal structure and the dominating magnetic interaction is AFM in nature. The detailed analysis of AC susceptibility (ACS), DC magnetization, memory effect and magnetic relaxation measurements confirm that these alloys exhibit SG behavior at low temperature.

Though Al is non-magnetic in nature, it plays a dominant role in dictating the magnetic properties of Mn-Fe-Al alloys by modifying the 3d band structure. With the increase in Al concentration, the Fe$_{0.89-x}$Mn$_{0.11}$Al$_x$ alloys undergo a structural transition from A2 to B2. The alloys with 30 at.\% $\leq$ x $\leq$ 40 at.\% adopt a B2 structure and are ordered antiferromagnetically \cite{ref10}. Similar magnetic behavior with a SG state is also reported for 42.5 at.\% of Al (Fe$_{27.5}$Mn$_{30}$Al$_{42.5}$) \cite{ref9}. The presence of competing magnetic interactions and atomic disorder is the source of glassy behavior in these systems. However, very recently, alloys with high Al concertation (Mn$_{30}$Fe$_{20}$Al$_{50}$) with a B2 (CsCl-type) structure and FM ground state are reported. The substitution of Cu at Fe sites in these alloys could allow an increase in the T$_C$ by 140 K \cite{ref20}.

Using Mn, Fe and Al, three kinds of stoichiometric FHAs can be designed: Mn$_2$FeAl, Fe$_2$MnAl, and Al$_2$MnFe. Physical properties of Mn$_2$FeAl \cite{ref17} and Fe$_2$MnAl \cite{ref12, ref13} have been investigated in detail. However, there is no report available on the structural and magnetic properties of the stoichiometric composition of the Al-rich alloy i.e., Al$_2$MnFe (Z$_2$XY-type). Herein, we explore the magnetic ground state, nature of magnetic exchange mechanism, itineracy of FM behavior and range of spin-spin interactions of single-phase polycrystalline Al$_2$MnFe alloy by ACS, critical magnetic behavior and magnetocaloric effect (MCE). Also, we examine the magnetocaloric performance of the alloy to unravel their potential for low-cost magnetic refrigeration applications.

The polycrystalline Al$_2$MnFe alloy crystallizes in a cubic crystal structure (space group Pm$\Bar{3}$m) and exhibits a soft FM behavior (T$_C$ = 122.9 K). The ACS analysis reveals a re-entrant SG behavior of the alloy at low temperature ($\sim$ 14 K). The comprehensive analysis of the critical isotherms using modified Arrott plots (MAPs) and Kouvel-Fisher plots (KFPs) univocally suggests that the underlying magnetic interactions are short range in nature and well explained using the 3D-Heisenberg universality class that is also verified using the renormalization group theory. An extensive MCE analysis suggests that the alloy undergoes a second order magnetic phase transition and also possesses a good magnetocaloric behavior as compared to other Mn-Fe-Al alloys.

\section{\label{sec:Synthesis and characterization techniques}Synthesis and characterization techniques}

Polycrystalline ingots of Al$_2$MnFe HA are synthesized by arc-melting technique using stoichiometric amounts of high purity Mn, Fe and Al ($\geq$ 99.95$\%$, Alfa Aesar) in the presence of argon gas. Before melting the constituent elements, a small ingot of titanium is melted to remove any oxygen-based impurity from the melting chamber. The ingots are remelted 3-4 times by inverting them each time. Since Mn has a propensity to vaporize, a 3 wt.$\%$ extra Mn is taken to prevent loss. The obtained ingots are then encapsulated in evacuated quartz tubes, annealed at 800 $^\circ$C for 7 days and finally quenched in ice water. The crystal structure of the samples is examined using X-ray diffraction (XRD), D8 - Discover (Bruker, Germany) (Cu-K$_\alpha$ source; $\lambda$ = 1.54 \AA). The XRD data is taken at room temperature ($2\theta$: 20$^\circ$ - 90$^\circ$, step size: 0.01$^\circ$). The composition of samples is investigated using energy dispersive X-ray spectroscopy (EDX) connected to a field emission scanning electron microscope (FE-SEM), JSM-7600F (JEOL, Japan). Temperature and magnetic field dependent magnetization measurements are carried out in a vibrating sample magnetometer (VSM) attached to a physical property measurement system (PPMS DynaCool - 9T, Quantum Design, U.S.A.). An AC susceptometer, attached to the same PPMS, is used to measure temperature dependent ACS at various frequencies.

\section{\label{sec:Results and discussion}Results and discussion}

\subsection{\label{sec:Structural and compositional analyses}Structural and compositional analyses}

\begin{figure}
    \includegraphics[width=0.45\textwidth]{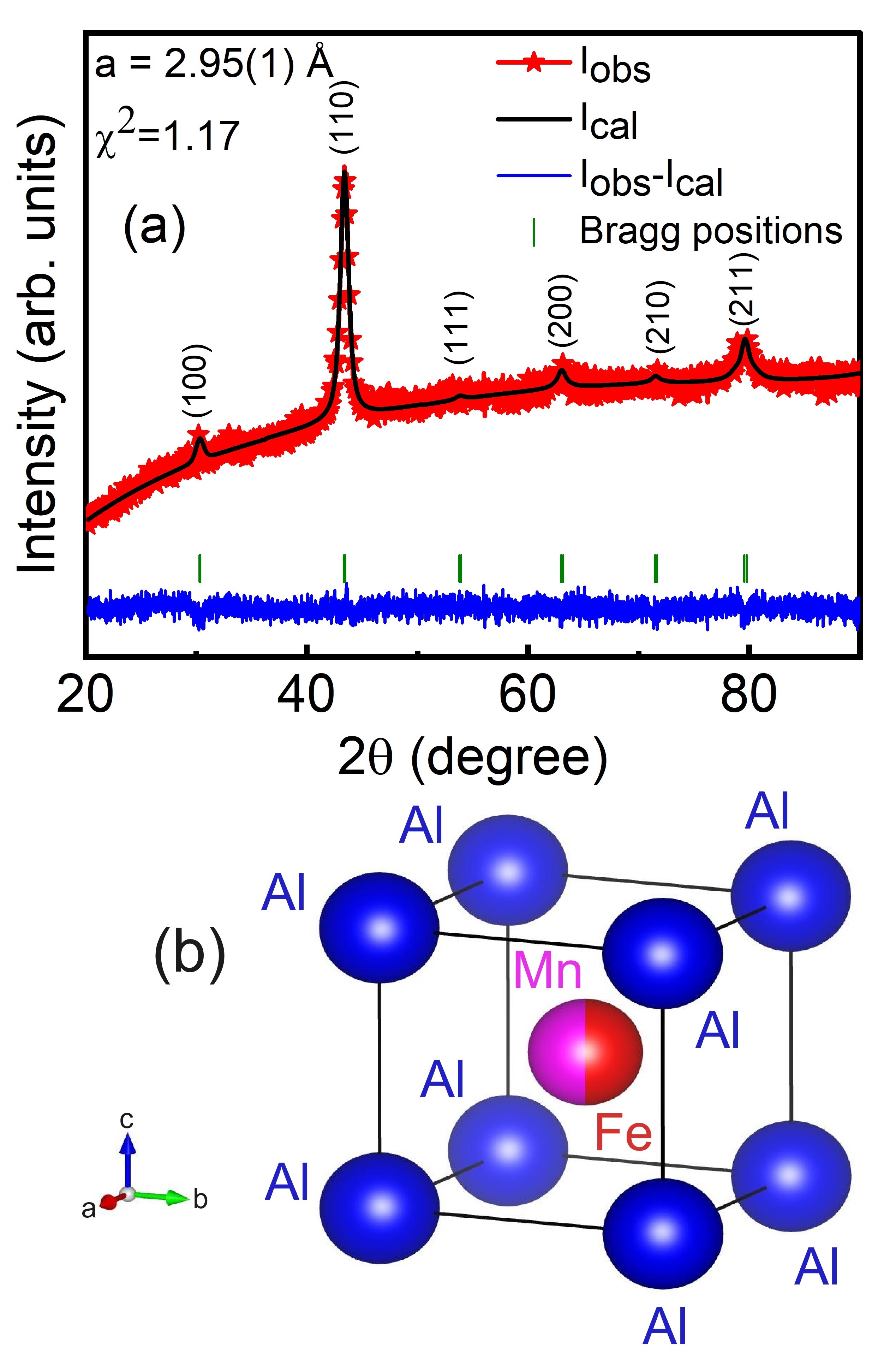}
    \caption{\label{fig:XRD}(a) Room temperature X-ray diffraction (XRD) pattern along with Rietveld refinement in the range 20$^\circ$ - 90$^\circ$ by choosing Pm$\Bar{3}$m space group. The red stars with line represent observed XRD pattern, black solid line represents calculated XRD pattern and solid blue line indicates the difference between the observed and calculated XRD patterns. The solid vertical green symbols indicate the Bragg’s positions. The value of goodness of fit ($\chi^2$ = 1.17) confirms that the observed and calculated XRD patterns are fitted well throughout the range. (b) Crystal structure of Al$_2$MnFe using the final Rietveld refined parameters. The blue spheres show Al atoms occupying 1a (0, 0, 0) Wyckoff positions and the red and pink sphere represents Mn and Fe atoms occupying 1b (1⁄2, 1⁄2, 1⁄2) Wyckoff positions.}
\end{figure}

Fig. \ref{fig:XRD}\hyperref[fig:XRD]{(a)} portrays the Rietveld refined room temperature powder X-ray diffraction (PXRD) data of the Al$_2$MnFe HA. The observed PXRD pattern is matched using the High Score Plus software. All the reflections present in the PXRD pattern are indexed using the body centered cubic (bcc) B2 phase (CsCl-type structure) with space group Pm$\Bar{3}$m, in which Al atoms occupy 1a (0, 0, 0) Wyckoff positions, whereas Mn and Fe atoms occupy 1b (1/2, 1/2, 1/2) Wyckoff positions as shown in Fig. \ref{fig:XRD}\hyperref[fig:XRD]{(b)}. The Rietveld analysis reveals that the studied alloy contains a single phase with lattice parameter a = 2.95(1) \AA~and unit cell volume V = 25.67(2)  \AA$^3$. Clearly, there is no additional peak associated with any impurity phase. The goodness of fit ($\chi^2$ = 1.17) provides a quantitative measure of the excellent match between the observed and the calculated XRD patterns. This is in excellent agreement with the earlier report on Mn$_{30}$Fe$_{20-x}$Cu$_{x}$Al$_{50}$ alloys \cite{ref20}.

Supplementary Fig. S1 displays the representative SEM image and EDS data of the prepared sample. The EDS data confirms the presence of constituents Mn, Fe and Al and suggests the absence of impurity elements. The average sample composition obtained from EDS is Al$_{50.22}$Mn$_{25.36}$Fe$_{24.56}$ which lies within $\pm$ 1 at.$\%$ of its nominal composition Al$_{50}$Mn$_{25}$Fe$_{25}$.

\subsection{\label{sec:Magnetic properties}Magnetic properties}

\subsubsection{\label{sec:Temperature and field dependent DC magnetization}Temperature and field dependent DC magnetization}

\begin{figure}
    \includegraphics[width=0.45\textwidth]{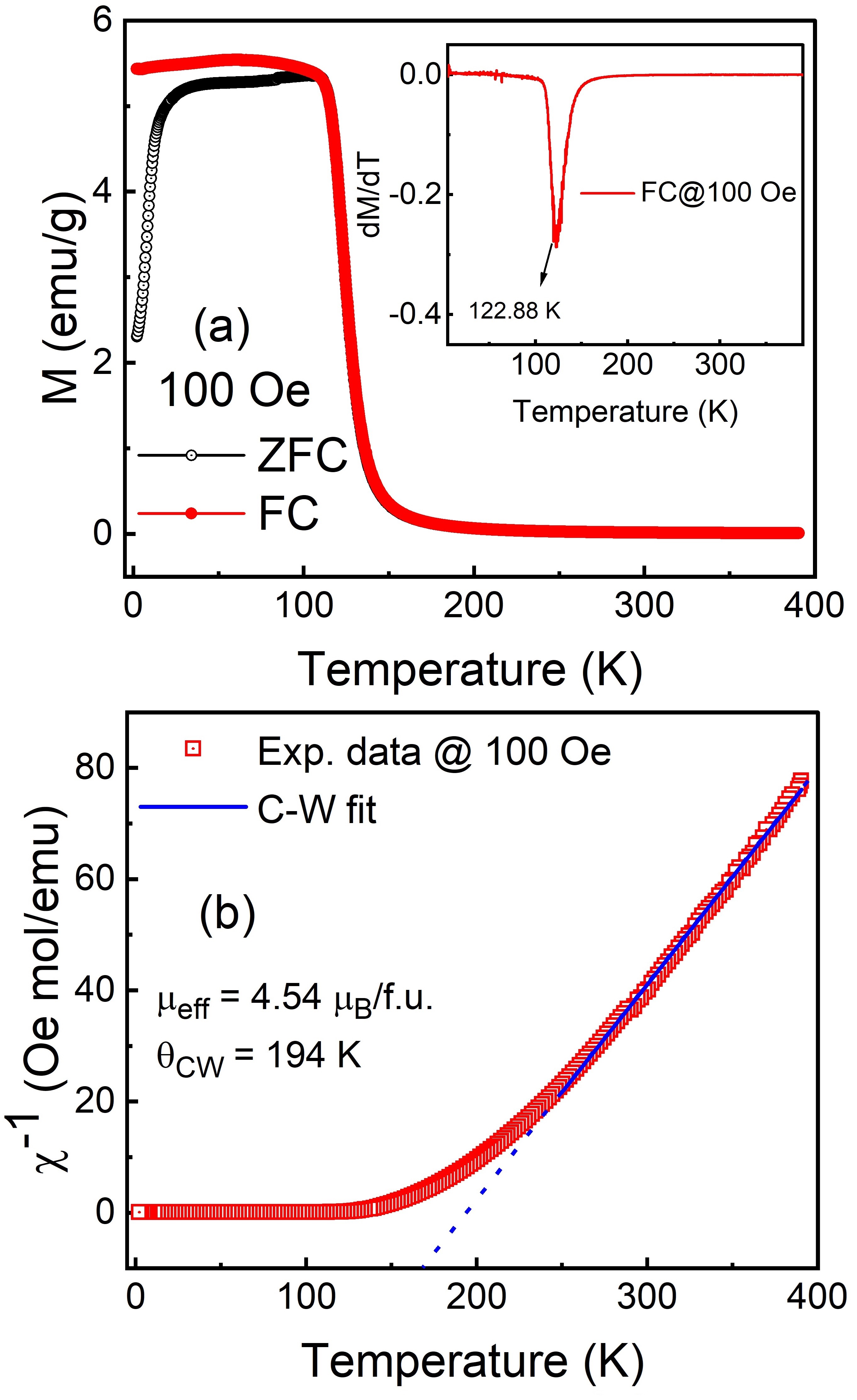}
    \caption{\label{fig:MT}(a) Temperature dependent magnetization in both ZFC (black dotted circle with solid line) and FC (red solid circle with line) modes in the range 2 K to 390 K in the presence of external applied magnetic field of 100 Oe. Inset displays the first derivative of magnetization as a function of temperature in FC mode which shows a sharp dip around 122.9 K, hinting at a magnetic transition. (b) Inverse DC susceptibility vs temperature along with Curie-Wiess fit (solid blue line) in the range 250 K - 390 K, extrapolated to the temperature axis.}
\end{figure}

Fig. \ref{fig:MT}\hyperref[fig:MT]{(a)} depicts the DC magnetization (M) as a function of temperature in an applied magnetic field (H=100 Oe) in zero-field-cooled (ZFC) (black dotted circle with solid line) and field-cooled (FC) (red solid circle with line) modes. For ZFC magnetization measurement (M$_{ZFC}$), the sample is cooled from 390 K to 2 K without any applied magnetic field. After stabilizing the temperature at 2 K, M$_{ZFC}$ is recorded from 2 K to 390 K under an applied magnetic field of 100 Oe during the warming cycle. For FC mode, magnetization (M$_{FC}$) is measured from 390 K to 2 K under the same applied magnetic field just after the M$_{ZFC}$ measurement. It is observed from Fig. \ref{fig:MT}\hyperref[fig:MT]{(a)} that, on decreasing the temperature from 390 K, magnetization increases rapidly around the critical temperature (T$_C$ = 122.9 K), which indicates the PM to FM phase transition. This transition is also confirmed through dM$_{FC}$(T)⁄dT vs T curve (inset of Fig. \ref{fig:MT}\hyperref[fig:MT]{(a)}). Interestingly, in low temperature region, a large bifurcation between M$_{ZFC}$ and M$_{FC}$ is observed which suggests the presence of another magnetic phase transition (MPT). The sharp drop in M$_{ZFC}$ observed on further lowering the temperature, is often attributed to the existence of a glassy magnetic phase or short-range correlations \cite{ref21}.

The inverse DC susceptibility ($\chi^{-1}\equiv$ H⁄M) as a function of temperature, recorded at an applied magnetic field of H = 100 Oe for Al$_2$MnFe is shown in Fig. \ref{fig:MT}\hyperref[fig:MT]{(b)} along with the Curie-Wiess (CW) fitting in the high temperature region (T $>$ T$_C$). The paramagnetic susceptibility can be described by CW law as:
\begin{equation}
    \chi (T) = \frac{C}{(T-\theta_{CW})},
    \label{eqn:1}
\end{equation}
where C is the Curie constant and $\theta_{CW}$ is the CW temperature. The fitted solid blue line (Fig. \ref{fig:MT}\hyperref[fig:MT]{(b)}) in the range 250 K to 390 K using equation (\ref{eqn:1}) yields the values of C = 2.58 emu K mol$^{-1}$ Oe$^{-1}$ and $\theta_{CW}$ = 194 K. The experimental effective magnetic moment is further determined to be $\mu_{eff}^{exp}$ = 4.54  $\mu_B$⁄f.u. using the relation $\mu_{eff}^{exp}=\sqrt{8C}$. The obtained positive value of $\theta_{CW}$ indicates that the FM interactions are dominant in the prepared alloy. The theoretical effective magnetic moment $\mu_{eff}^{th}=\sqrt{(\mu_{Mn})^2+(\mu_{Fe})^2}$ = 7.68  $\mu_B$⁄f.u.  (using magnetic moments of Mn and Fe to be 5.92 $\mu_B$ and 4.90 $\mu_B$, respectively) is higher than the experimental one. The difference between experimental and theoretical magnetic moments may be attributed to the presence of short-range AFM correlations in the PM region.
\begin{figure}
    \includegraphics[width=0.45\textwidth]{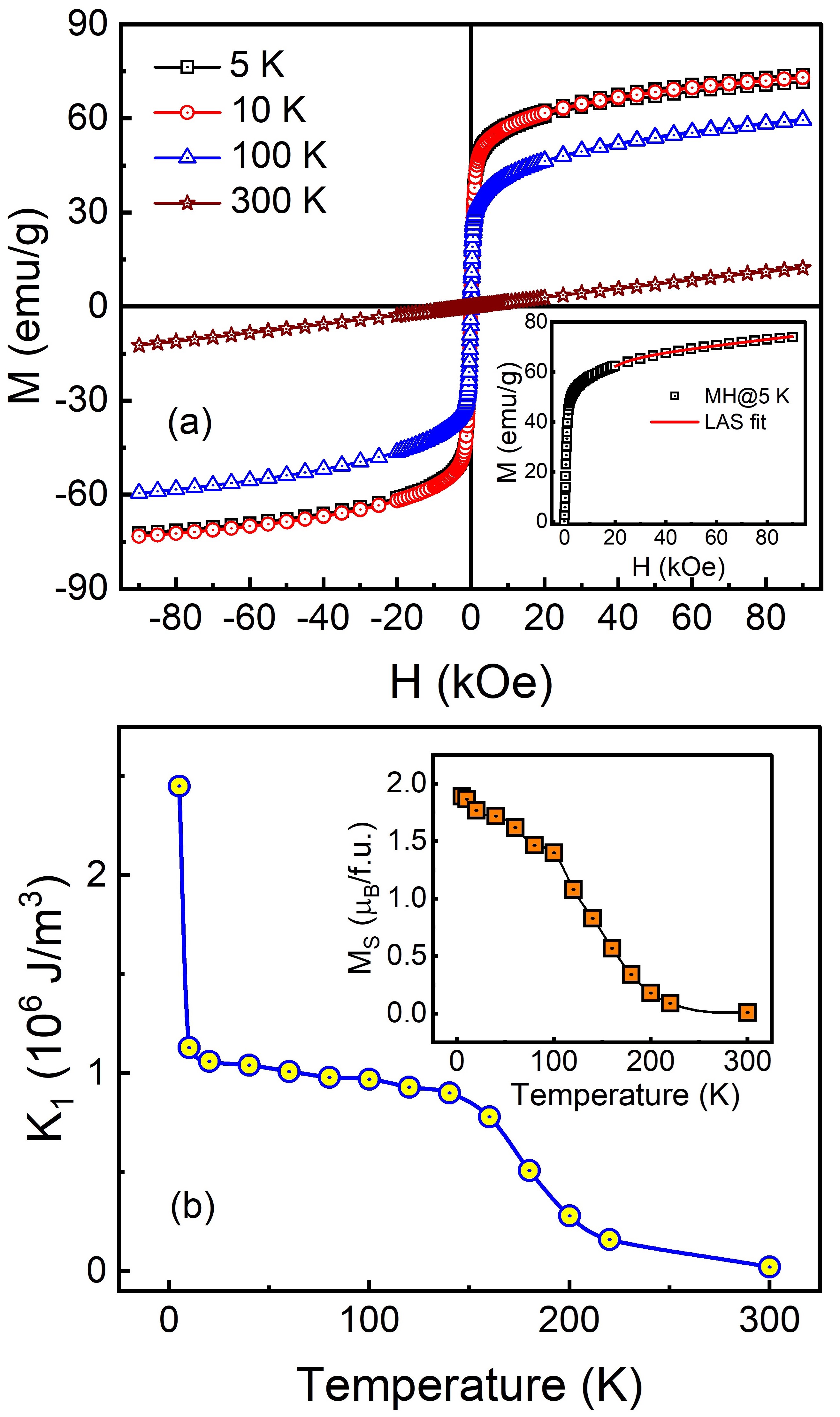}
    \caption{\label{fig:MH}(a) Isothermal magnetization as a function of applied magnetic field in the range $\pm$ 90 kOe at selected temperatures. Inset represents the virgin MH curve at 5 K along with the fit to the equation of law of approach to saturation (solid red line). (b) The temperature variation of cubic magneto-crystalline anisotropy K$_1$ (solid blue line: guide to eye). Inset represents the saturation magnetization M$_S$ as a function of temperature (solid black line: guide to eye).}
\end{figure}

The M(H) hysteresis loops (Fig. \ref{fig:MH}\hyperref[fig:MH]{(a)}), recorded at various temperatures, suggest the soft FM behavior of the material. However, unlike a typical ferromagnet, even with an applied field as high as 90 kOe, saturation in magnetization is not observed. This may be indicative of the presence of AFM interactions \cite{ref22}. The maximum magnetic moment ($\mu_S$) measured at 5 K with 90 kOe is found to be 2.18  $\mu_B$⁄f.u. The values of coercivity (H$_C$) and remanence (M$_R$) at 5 K are 0.10 kOe and 0.17  $\mu_B$⁄f.u., respectively which decrease with increasing temperature, as expected for a FM system.

Additionally, the magnetic moment derived from Curie constant and the maximum magnetic moment measured at lowest temperature ($\mu_S$) are used to calculate the Rhodes-Wohlfarth ratio, RWR = $\mu_C$⁄$\mu_S$, where $\mu_C=\sqrt{(\mu_{eff}^{exp})^2+1}-1$. RWR is used to determine the nature of interactions in a FM system. If RWR is 1, the dominant interactions are localized. For itinerant interactions the RWR is greater than 1 \cite{ref23, ref24}. RWR value for the prepared alloy is $\sim$ 1.67, which indicates the itinerant nature of underlying exchange interactions.

To gain a deeper understanding of the nature of itinerant exchange interactions and the degree of itineracy in the studied Al$_2$MnFe alloy, the MH isotherm at T$_C$ is analyzed on the basis of Self-Consistent Renormalizaton (SCR) theory which provides a comprehensive framework for understanding itinerant ferromagnets \cite{ref25, ref26, ref27}. According to this theory, the MH isotherm at T$_C$ follows the given relation:
\begin{equation}
    M^4=\frac{1}{4.671}\left(\frac{T_C^2}{T_A^3}\right)\left(\frac{H}{M}\right),
    \label{eqn:2}
\end{equation}
where T$_A$ represents the dispersion of spin fluctuation spectrum in the wave vector space whose value can be estimated from the slope of the linear fit of equation (\ref{eqn:2}). The value of slope is obtained to be 1.31 $\times$ 10$^{-4}$ ($\mu_B/f.u.)^5$/Oe (see supplementary Fig. S2), which yields T$_A$ = 286.5 K. The SCR theory suggests that the T$_C$ of the system is related to T$_A$ by the following relation:
\begin{equation}
    T_C=(60c)^{-3/4}P_S^{3/2}T_A^{3/4}T_0^{1/4},
    \label{eqn:3}
\end{equation}
where c = 0.3353 is a constant, P$_S$ is the spontaneous magnetization of the system in its ground state, and T$_0$ is the dispersion of spin fluctuation spectrum in the energy space. From the values of P$_S$, T$_C$, and the obtained value of T$_A$, T$_0$ is calculated to be 1525 K. According to the SCR theory, the ratio $\frac{T_C}{T_0}$ determines the degree of itineracy or localization of magnetic moments in the system. For $\frac{T_C}{T_0}\sim 1$, the system is dominated by localized magnetic moments, whereas $\frac{T_C}{T_0}<<1$ suggests a high degree of itineracy of magnetic moments and hence a strong itinerant character of the electrons in the system. In Al$_2$MnFe alloy, $\frac{T_C}{T_0}=0.078$ indicates a strong itinerant nature of exchange interactions which is in excellent agreement with RWR discussed previously \cite{ref27}.

Furthermore, the high field region of hysteresis loop may be used for estimating the magneto-crystalline anisotropy of uniaxial and cubic crystals using the law of approach to saturation (LAS) \cite{ref28}. Typically, in the high field region where the ferromagnet's magnetization nearly saturates, the magnetization along the field direction is given by \cite{ref29, ref30}:
\begin{equation}
    M=M_S\left ( 1-\frac{b}{H^2}\right)+\chi_h H,
    \label{eqn:4}
\end{equation}
where M$_S$ is the saturation magnetization, b is associated with magnetic anisotropy, given by $b=\frac{8}{105}\frac{K_1^2}{M_S^2}$; K$_1$ is the cubic magneto-crystalline anisotropy and $\chi_h$ is the susceptibility at high field. The inset of Fig. \ref{fig:MH}\hyperref[fig:MH]{(a)} depicts the representative MH virgin curve at 5 K along with fit to equation (\ref{eqn:4}) in the high field region (20 kOe $\leq$ H $\leq$ 90 kOe). The obtained values of b, M$_S$ and $\chi_h$ are 2.60 $\times$ 10$^7$ Oe$^2$, 1.89 $\mu_B$, and 1.15 $\times$ 10$^{-4}$ emu/gOe, respectively. The temperature dependent K$_1$ and M$_S$ are depicted in Fig. \ref{fig:MH}\hyperref[fig:MH]{(b)}. Temperature has a significant impact on K$_1$ and its value decreases with increasing temperature that may be attributed to a decrease in interparticle interactions caused by thermal fluctuations \cite{ref31}. At 5 K (1.18 $\times$ 10$^6$ J⁄m$^3$), the value of K$_1$ is two orders of magnitude higher than its value at room temperature (2.15 $\times$ 10$^4$ J⁄m$^3$). Similar trend of K$_1$ is observed for other FM materials (Rh$_2$CoSb \cite{ref32}, MnFe$_2$O$_4$ \cite{ref33}, La$_{1.36}$Sr$_{1.64}$Mn$_2$O$_7$ \cite{ref34}).

\subsubsection{\label{sec:Re-entrant spin-glass behavior}Re-entrant spin-glass behavior}

\begin{figure*}
    \includegraphics[width=0.9\textwidth]{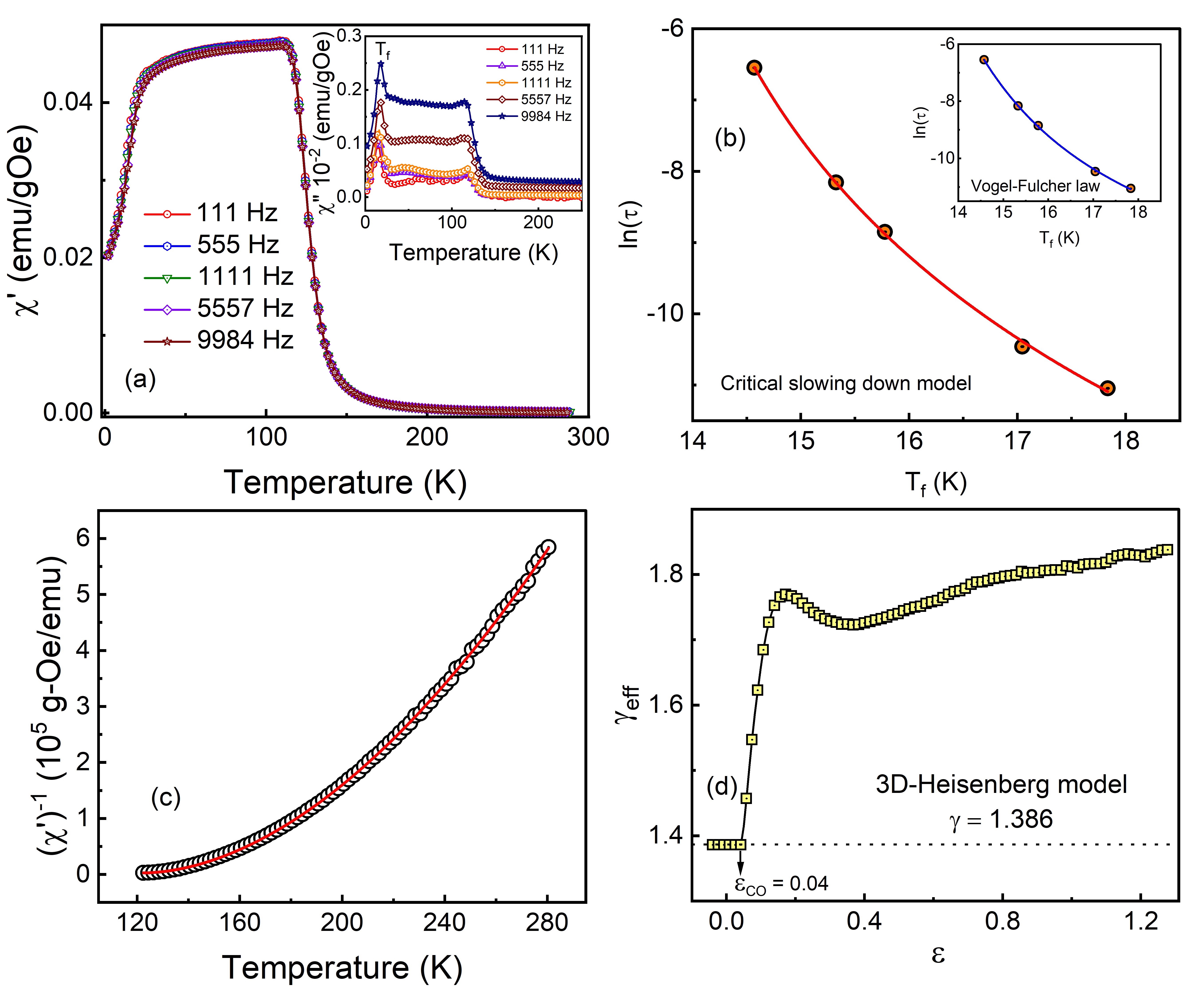}
    \caption{\label{fig:ACS}(a) The real part of ACS ($\chi'$) as a function of temperature at various frequencies. Inset represents the temperature dependent imaginary part of ACS ($\chi''$) at various frequencies which shows a peak around the SG freezing temperature (T$_f$). (b) $\ln⁡(\tau)$  vs T$_f$ with fitting to critical slowing down model (solid red line) and Vogel-Fulcher law (Inset) (solid blue line). (c) Temperature dependent inverse of the real part of ACS of the prepared alloy. The solid red line represents the range-of-fit to equation (\ref{eqn:10}). (d) The critical exponent ($\gamma_{eff}$) as a function of reduced temperature above T $\geq$ T$_C$ which indicates that the studied alloy belongs to the 3D-Heisenberg model in the critical asymptotic region.}
\end{figure*}

The appearance of an abrupt downturn in M$_{ZFC}$ at low temperature indicates the existence of a glassy phase. To understand the genesis of the magnetic transition at low temperature region, temperature dependent ACS is measured at various frequencies with an excitation field of 10 Oe. The real part of ACS ($\chi'$) at various frequencies as a function of temperature is presented in Fig. \ref{fig:ACS}\hyperref[fig:ACS]{(a)} which shows similar feature as M$_{ZFC}$. The temperature dependence of the imaginary part of ACS ($\chi''$) at various frequencies is displayed in the inset of Fig. \ref{fig:ACS}\hyperref[fig:ACS]{(a)}. A sharp peak around T$_f$ = 13.94 K in $\chi''$ vs T, corresponding to the SG freezing temperature, is observed. The T$_f$ shifts to higher temperature with increasing frequency (f$_1$ = 111 Hz to f$_5$ = 9984 Hz) which is a typical signature of glassy systems: spin-glass (SG)/non-interacting superparamagnetic (SPM) systems \cite{ref35}.

Mydosh parameter, the relative shift in the freezing temperature T$_f$ per decade of frequency, allows to distinguish between SG and SPM systems \cite{ref35}. Mydosh parameter is expressed as:
\begin{equation}
    \delta_M=\frac{\Delta T_f}{T_f \Delta(log_{10} f)},
    \label{eqn:5}
\end{equation}
where the frequency dependent relative shift in T$_f$ is expressed as $\Delta T_f=T_{f_5}-T_{f_1}$ and $\Delta (log_{10} f)=(log_{10}(f_5))-(log_{10} (f_1))$. For SG, $\delta_M$ varies between 0.005 and 0.09, whereas for SPM systems, it is typically greater than 0.1 \cite{ref35, ref36}. For Al$_2$MnFe, the obtained value of $\delta_M$ = 0.11 confirms its cluster SG nature similar to Fe$_2$CrAl \cite{ref60}.

To gain further insight into SG systems, the frequency dependence of T$_f$ is analysed using the critical slowing down model according to the dynamic scaling theory \cite{ref35, ref36, ref37}:
\begin{equation}
    \tau =\tau_0  \left ( \frac{T_f}{T_g} -1\right)^{-z\nu},
    \label{eqn:6}
\end{equation}
where the observed frequency $f$ is accompanied by a characteristic relaxation time denoted by $\tau=\frac{1}{2\pi f}$, the relaxation time for flipping a single spin is $\tau_0$, the SG ordering temperature in the limit $f \rightarrow 0$ is T$_g$, and the dynamical exponent is $z\nu$. The range of $\tau_0$ enables to distinguish various glassy systems. To fit the frequency dependent peak temperatures (T$_f$), equation (\ref{eqn:6}) is reformulated as:
\begin{equation}
    \ln⁡(\tau)=\ln⁡(\tau_0)-z\nu\ln⁡\left(\frac{T_f}{T_g}-1\right).
    \label{eqn:7}
\end{equation}

A least square fit of $\ln⁡(\tau)$ vs T$_f$ plot using equation (\ref{eqn:7}) is depicted in Fig. \ref{fig:ACS}\hyperref[fig:ACS]{(b)}. The obtained values of $\tau_0$, T$_g$ and $z\nu$ are 2.94(1) $\times$ 10$^{-7}$ s, 13.13(2) K and 3.8(3), respectively, confirming the SG nature of the system. $\tau_0$ and $z\nu$ provide a deeper understanding of the SG dynamics. SG systems have $z\nu$ and $\tau_0$ in the range of 4 - 12 and 10$^{-13}$ - 10$^{-4}$ s, respectively. The value of $z\nu$ is slightly less than 4, similar to that reported for Pr$_{117}$Co$_{54.5}$Sn$_{115.2}$ \cite{ref38} and Mn$_5$Sn$_3$ \cite{ref39} cluster SG systems. Further $\tau_0$ lies between 10$^{-8}$ - 10$^{-4}$ s for cluster SGs, in which relaxation is slower than conventional SGs because there are more spins involved in each relaxation due to the presence of spin clusters \cite{ref35, ref37}. Therefore, the measured values of $z\nu$ and $\tau_0$ confirm that the prepared alloy belongs to the family of cluster SGs.

The phenomenological Vogel-Fulcher (VF) law \cite{ref35, ref40}, which considers the interaction between the spins, is another dynamical scaling law used to study glassy dynamics. This law describes the frequency dependent T$_f$ as:
\begin{equation}
    \tau=\tau_0~exp\left(\frac{E_a}{k_B(T_f-T_0)}\right),
    \label{eqn:8}
\end{equation}
where T$_0$ is the Vogel-Fulcher freezing temperature representing the interaction strength between interacting spin clusters, k$_B$ is the Boltzmann’s constant, and E$_a$ is the activation energy. Equation (\ref{eqn:8}) is reformulated as:
\begin{equation}
    \ln⁡(\tau) = \ln⁡(\tau_0) + \left(\frac{E_a}{k_B (T_f-T_0)}\right).
    \label{eqn:9}
\end{equation}

A least square fit of $\ln⁡(\tau)$ vs T$_f$ to equation (\ref{eqn:9}) is depicted in the inset of Fig. \ref{fig:ACS}\hyperref[fig:ACS]{(b)}. The obtained values of  T$_0$ and (E$_a$⁄k$_B$) are 10.6(4) K and 39.7(6), respectively, that are similar to the values reported for other cluster SG systems \cite{ref38, ref39}. The ratio (E$_a$⁄k$_B$T$_0$) for SGs should be less than 3 \cite{ref17}, and our calculated value for the studied alloy is 1.43, further verifying its SG nature. Similar values of parameter (E$_a$⁄k$_B$T$_0$) are reported for other SG systems such as Mn$_2$FeAl (1.9) \cite{ref17}, Mn$_{1.5}$Fe$_{1.5}$Al (1.9) \cite{ref18}, IrMnGa (2) \cite{ref41} and FeRuMnGa (2.6) \cite{ref42}.

\subsubsection{\label{sec:Critical analysis}Critical analysis}

Critical analysis near the FM-PM phase transition temperature is commonly employed to explore the nature of magnetic interactions in solids. These interactions are generally described using various microscopic models such as Mean field, 3D-Heisenberg, 3D-Ising and 3D-XY. The analysis of $\chi'$ vs T data is one of the methods to identify the most appropriate model that describes the underlying magnetic interactions.

The intrinsic inverse susceptibility ($\chi^{-1}$ (T)) is related to the measured real part of ACS ($\chi'$(T)) by the following expression \cite{ref43, ref44}:
\begin{equation}
    (\chi'(T))^{-1} = \chi^{-1}(T)+ 4\pi N.
    \label{eqn:10}
\end{equation}

$\chi$(T) diverges at (T = T$_C$), and hence equation (\ref{eqn:10}) becomes $(\chi'(T))^{-1}=4\pi N$, where N is the demagnetization factor. In the paramagnetic state (T $\geq$ T$_C$), $\chi^{-1}$ (T) varies with temperature as:
\begin{equation}
    \chi^{-1}(T) = A_{eff} (\epsilon)^{\gamma_{eff}}; (\epsilon>0),
    \label{eqn:11}
\end{equation}
where $\epsilon=((T-T_C)/T_C)$ is reduced temperature, A$_{eff}$ is effective critical amplitude and $\gamma_{eff}$ is effective critical exponent. The thermal variation of measured ACS is fitted using the expression $(\chi'(T))^{-1}=A_{eff}(\epsilon)^{\gamma_{eff}}+4\pi N$, employing the ``range-of-fit" analysis \cite{ref44} to extract the values of $\gamma_{eff}$.  $(\chi'(T))^{-1}$ along with the fit to the above expression is shown in Fig. \ref{fig:ACS}\hyperref[fig:ACS]{(c)} and the corresponding temperature dependence of $\gamma_{eff}$ is displayed in Fig. \ref{fig:ACS}\hyperref[fig:ACS]{(d)}.

In the asymptotic critical region ($\epsilon$ = 0 or T = T$_C$), A$_{eff}$ = A and $\gamma_{eff}=\gamma$ in equation (\ref{eqn:11}) represent the true critical behavior. From Fig. \ref{fig:ACS}\hyperref[fig:ACS]{(d)}, it is clear that as $\epsilon\rightarrow0$, $\gamma_{eff}\rightarrow \gamma$ = 1.386, which is the critical exponent associated with the 3D-Heisenberg model. As the temperature increases gradually, $\gamma_{eff}$ continuously increases after a crossover temperature ($\epsilon_{CO}$ = 0.04). The above analysis indicates that the system exhibits isotropic short-range 3D-Heisenberg-type behavior in the asymptotic critical region.

\begin{figure*}
    \includegraphics[width=0.95\textwidth]{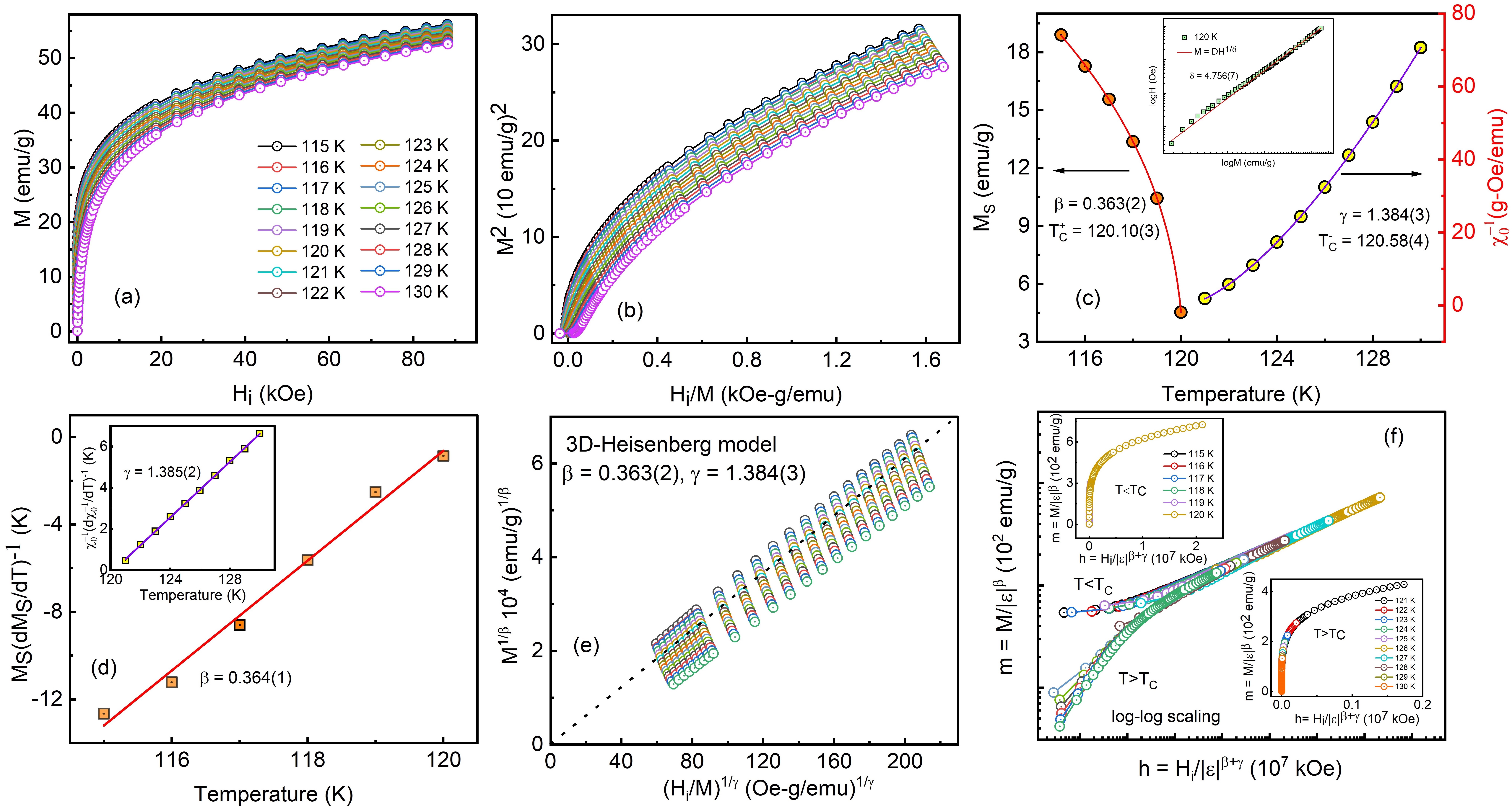}
    \caption{\label{fig:CA}(a) Isothermal magnetization as a function of applied magnetic field up to 90 kOe at selected temperatures in the temperature range 115 K - 130 K with $\Delta$T = 1 K. (b) Arrott plots: M$^2$ as a function of H$_i$/M for recorded isotherms. (c) Saturation magnetization (M$_S$) and inverse initial susceptibility ($\chi_0^{-1}$) as functions of temperature obtained from the MAP along with the power law fitting using equations (\ref{eqn:13}) and (\ref{eqn:14}). Inset represents the critical isotherm at 120 K fit using equation (\ref{eqn:15}) and shown in logarithmic scale. (d) Kouvel-Fisher plots: Temperature dependent M$_S$ (dM$_S$/dT)$^{-1}$ and $\chi_0^{-1}$ ($d\chi_0^{-1}$/dT)$^{-1}$ along with their linear fit. (e) Final Modified Arrott plots generated using the final critical exponent obtained from the fitting of M$_S$ and $\chi_0^{-1}$. (f) The renormalized magnetization (m) as a function of renormalized magnetic field (h) based on the scaling equation (\ref{eqn:18}) in the logarithmic scale. Inset shows the same plot in normal scale.}
\end{figure*}

To have an in-depth insight into the critical behavior, MH isotherms are recorded around T$_C$  (115 K $\leq$ T $\leq$ 130 K; $\Delta$T = 1 K). The MH isotherms in the critical region are presented in Fig. \ref{fig:CA}\hyperref[fig:CA]{(a)}. It is observed that the magnetization rapidly increases in the low field region ($\leq$ 7 kOe) and thereafter increases gradually. To obtain valuable information about the characteristics of MPT, the corresponding Arrott plots ([M(T,H)]$^2$ vs H$_i$⁄M(T,H)) are analysed and presented in Fig. \ref{fig:CA}\hyperref[fig:CA]{(b)}. The observed positive slopes in the Arrott plots affirm the second order magnetic phase transition (SO-MPT) according to the Banerjee criterion \cite{ref45}. However, non-linear behavior of the Arrott plots at high fields and the critical isotherm not passing through the origin indicate that the mean-field model ($\beta$ = 0.5, $\gamma$ = 1.0) is not applicable in the present case. Therefore, to further investigate about the appropriate universality class, Arrott plots are modified using Arrott-Noakes equation expressed as \cite{ref46}:
\begin{equation}
    \left(\frac{H}{M}\right)^{1\slash{\gamma}} = a\epsilon + bM^{1\slash{\beta}},
    \label{eqn:12}
\end{equation}
where a and b are fitting parameters, and $\gamma$ and $\beta$ are the critical exponents. The modified MAPs are constructed using various theoretical critical exponents predicted for three dimensional (3D) systems such as 3D-XY ($\beta$ = 0.346, $\gamma$ = 1.316), 3D-Heisenberg ($\beta$ = 0.365, $\gamma$ = 1.386) and 3D-Ising ($\beta$ = 0.325, $\gamma$ = 1.241) models (see supplementary Fig. S3(a-c)). Using the correct choice of $\beta$ and $\gamma$, isotherms can be constructed around T$_C$ which are straight and parallel to each other and a linear extrapolation of the isotherm at T$_C$ from high field passes through the origin. In order to choose the best suitable model, the relative slope RS = S(T)⁄S(T$_C$) is plotted as a function of temperature for each model (supplementary Fig. S3(d)). For an ideal model, the value of RS is close to 1. It is observed that the 3D-Heisenberg model has a relatively smaller deviation compared to other models. Therefore, the MAPs for 3D-Heisenberg model are employed to further refine $\beta$ and $\gamma$ by an iterative method using the following power law dependent magnetization (M$_S$(T)) and inverse susceptibility ($\chi_0^{-1}$(T)) below and above T$_C$, respectively.
\begin{equation}
    M_S (T) = M_0 \vert \epsilon\vert ^{\beta}; T<T_C,
    \label{eqn:13}
\end{equation}
\begin{equation}
    \chi_0^{-1}(T) = (h_0/M_0 ) \vert\epsilon\vert^{\gamma}; T>T_C,
    \label{eqn:14}
\end{equation}
where M$_0$ and (h$_0$/M$_0$) are the critical amplitudes related to M$_S$ and $\chi_0^{-1}$, respectively. The linear fit of MAPs (M$^{1\slash{\beta}}$  vs (H⁄M)$^{1\slash{\gamma}}$)  using $\beta$ = 0.365, $\gamma$ = 1.386 in the high field region provides intercepts on the y-axis for T $<$ T$_C$ and on the x-axis for T $>$ T$_C$. These y- and x-intercepts allow to extract the values of M$_S$(T) and $\chi_0^{-1}$(T), respectively.  The obtained values of M$_S$(T) and $\chi_0^{-1}$(T) are fitted using equations (\ref{eqn:13}) and (\ref{eqn:14}) to determine the refined values of $\beta$ and $\gamma$. The new values of $\beta$, $\gamma$ and T$_C$ are further used to construct the new sets of MAPs. The above procedure is repeated until $\beta$ and $\gamma$ are convergent. Thus, the final fit (shown in Fig. \ref{fig:CA}\hyperref[fig:CA]{(c)}) yields the values of $\beta$ = 0.363(2) with T$_C$ = T$_C^-$ = 120.58(4) K and $\gamma$ = 1.384(3) with  T$_C$ = T$_C^+$ = 120.10(3) K. $\gamma$ and $\beta$ are related to another critical exponent $\delta$ via Widom relation $\delta$ = 1 + $\gamma$/$\beta$. Using the refined values of $\beta$ and $\gamma$, the obtained value of $\delta$ is 4.81 which is in excellent agreement with the value of  $\delta$ = 4.8 for the 3D-Heisenberg universality class. Moreover, the magnetization and applied magnetic field at the critical temperature (T = T$_C$) are related through $\delta$ by the following equation:
\begin{equation}
    M = D(H)^{1/{\delta}}; T=T_C,
    \label{eqn:15}
\end{equation}
where D is the critical amplitude related to M(H, T = T$_C$). The critical MH isotherm (120 K), fitted by equation (\ref{eqn:15}), gives the value of $\delta$ $\sim$ 4.756(7) (shown in the inset of Fig. \ref{fig:CA}\hyperref[fig:CA]{(c)}) which is close to the value of $\delta$ obtained from the Widom relation. The obtained parameters ($\beta$, $\gamma$ and T$_C$) from the analysis of MAPs can be further verified using Kouvel-Fisher plots (KFPs) \cite{ref46}:
\begin{equation}
    M_S \left(dM_S/dT\right)^{-1}=\frac{(T_C-T)}{\beta},
    \label{eqn:16}
\end{equation}
\begin{equation}
    \chi_0^{-1} (d\chi_0^{-1}/dT)^{-1}=\frac{(T-T_C)}{\gamma}.
    \label{eqn:17}
\end{equation}
\begin{table*}
    \caption{\label{tab:CA}The critical exponents of Al$_2$MnFe HA compared with other works on Mn-based materials.}
    \begin{ruledtabular}
        \begin{tabular}{ccccccc}
            \textbf{Sample/model} & \textbf{Technique} & \textbf{T$_C$ (K)} & \bm{$\beta$} & \bm{$\gamma$} & \bm{$\delta$} & \textbf{Reference}\\
            \hline
            Mean field & Theory & - & 0.5 & 1 & 3 & \cite{ref43}\\
            3D-Heisenberg & Theory & - & 0.365 & 1.386 & 4.80 & \cite{ref43}\\
            3D-Ising & Theory & - & 0.325 & 1.241 & 4.82 & \cite{ref43}\\
            3D-XY & Theory & - & 0.346 & 1.316 & 4.81 & \cite{ref43}\\
            Mn$_{30}$Fe$_{12}$Cu$_8$Al$_{50}$ & MAP & 301.86 & 0.371 & 1.326 & 4.51 & \cite{ref20}\\
            & KF & 301.12 & 0.386 & 1.273 & 4.30 & \cite{ref20}\\
            Fe$_{1.25}$Mn$_{0.75}$CrAl & MAP & 26.86 & 0.347 & 1.6 & 5.61 & \cite{ref21}\\
            & KF & 26.79 & 0.345 & 1.61 & 5.60 & \cite{ref21}\\
            Al$_2$MnFe & MAP & 120.10(3) & 0.363(2) & 1.384(3) & 4.81 & This work\\
            & KF & 120.20(1) & 0.37(1) & 1.40(2) & 4.78 & This work\\
            & Critical Isotherm & - & - & - & 4.75 & This work\\
        \end{tabular}
    \end{ruledtabular}
\end{table*}

Fig. \ref{fig:CA}\hyperref[fig:CA]{(d)} and its inset represent the temperature dependent M$_S$ (dM$_S$/dT)$^{-1}$ and $\chi_0^{-1}$ ($d\chi_0^{-1}$/dT)$^{-1}$ which demonstrate a linear relation with slope 1/$\beta$ and 1/$\gamma$, respectively. The linear fits of the KFPs provide the transition temperature and critical exponents as $\beta$ = 0.37(1) and $\gamma$ = 1.40(2), which are consistent with the values previously obtained from the MAP analysis. The obtained values of critical exponents and T$_C$ are tabulated in Table \ref{tab:CA}. By using the refined values of $\beta$ and $\gamma$ the MAPs are constructed and displayed in Fig. \ref{fig:CA}\hyperref[fig:CA]{(e)}. As expected, these MAPs feature parallel and straight isotherms in the high field regime and the critical isotherm passes through the origin. The obtained values of critical exponents $\beta$ and $\gamma$ using MAPs and KFPs represent the true critical exponents that are in good agreement with those of the 3D-Heisenberg model. To further validate the consistency of the critical exponents with the corresponding model, it is necessary to justify these exponents through scaling hypothesis as follows \cite{ref46, ref47}:
\begin{equation}
    M(H,\epsilon) \vert \epsilon \vert ^{-\beta}=f_\pm(H\vert \epsilon \vert^{-(\beta +\gamma)}),
    \label{eqn:18}
\end{equation}   
where f$_+$ (for T $>$ T$_C$) and f$_-$ (for T $<$ T$_C$) are the regular analytical functions. The above equation can be formulated as m = f$_\pm$(h), where m $\equiv$ M(H,$\epsilon$)$\vert\epsilon\vert^{-\beta}$ and h $\equiv$ H$\vert \epsilon\vert^{-(\beta + \gamma)}$ are rescaled magnetization and rescaled field, respectively. The m(h) plots and the logarithmic scaled curve are shown in Fig. \ref{fig:CA}\hyperref[fig:CA]{(f)}. The m(h) curves collapse into two distinct branches below and above T$_C$, indicating that the critical exponents are correctly chosen and thus confirming the validity of the scaling relation. This analysis further validates the reliability and consistency of obtained critical exponents with scaling theory.

The detailed analysis of intrinsic ACS, critical analysis of MH isotherms using MAPs, KFPs and scaling theory conclusively suggest that the Al$_2$MnFe HA belongs to the 3D-Heisenberg universality class and that the underlying magnetic behavior is dominated by isotropic short-range 3D-Heisenberg exchange interactions.

Furthermore, it is important to understand the nature and range of the underlying magnetic interactions in itinerant ferromagnets and this can be interpreted from the re-normalization group theory \cite{ref48}. According to this theory, the exchange interactions among the spins of itinerant electrons follow J(r) $\approx$ r$^{-(d+\sigma)}$, where r is the distance, d is the dimensionality of the system, and $\sigma$ represents the range of exchange interactions and is a positive constant \cite{ref47, ref49}. The range of interactions can be classified into three categories. For $\sigma \leq 1.5$, the exchange interactions are long-range mean-field type, for $\sigma \geq 2$, the interactions follow short-range 3D-Heisenberg-type, and for 1.5 $< \sigma <$ 2, the exchange interactions are of an intermediate range \cite{ref47, ref50}. The value of $\sigma$ for the studied system is estimated from equation (\ref{eqn:19}) \cite{ref47, ref49, ref51}, where n represents the spin dimensionality of the system, $\Delta \sigma = \sigma - d/2$ and $G(\frac{d}{2})= 3-(1/4)(d/2)^2$. Considering the values of d = n = 3 (as inferred from the MAP, KF and critical isotherm analyses), the value of $\sigma$ is optimized in such a way that equation (\ref{eqn:19}) yields $\gamma=1.384$ which is the value of $\gamma$ obtained from the MAP analysis. This optimized value of $\sigma$ is found to be 1.936, which denotes that the exchange interactions present in the studied system deviates from a truly short-range isotropic 3D-Heisenberg-type of interactions ($\sigma\geq$ 2) \cite{ref47}. The critical exponents corresponding to the correlation length ($\nu$) and specific heat ($\alpha$) can be further calculated using the relations $\nu= \gamma/\sigma$ and $\alpha=2-2\beta-\gamma$. The obtained values of $\nu =0.715$ and $\alpha = -0.11$ are in excellent agreement with the values corresponding to a 3D-Heisenberg system ($\nu =0.693$ and $\alpha = -0.115$) \cite{ref47, ref52}. To summarize, the SCR theory suggests that the studied system slightly deviates from a truly short-range type of exchange interactions (as in the case of 3D-Heisenberg systems) and possibly hosts intermediate to short-range exchange interactions within the 3D-Heisenberg universality class.
\begin{widetext}
    \begin{equation}
        \gamma = 1+ \frac{4(n+2)}{d(n+8)} \Delta \sigma + \frac{8(n+2)(n-4)}{d^2(n+8)^2}\left[1+\frac{2G(\frac{d}{2})(7n+20)}{(n-4)(n+8)}\right]\Delta {\sigma}^2
        \label{eqn:19}
    \end{equation}
\end{widetext}

\subsubsection{\label{sec:Magnetocaloric study}Magnetocaloric study}

\begin{figure*}
    \includegraphics[width=0.88\textwidth]{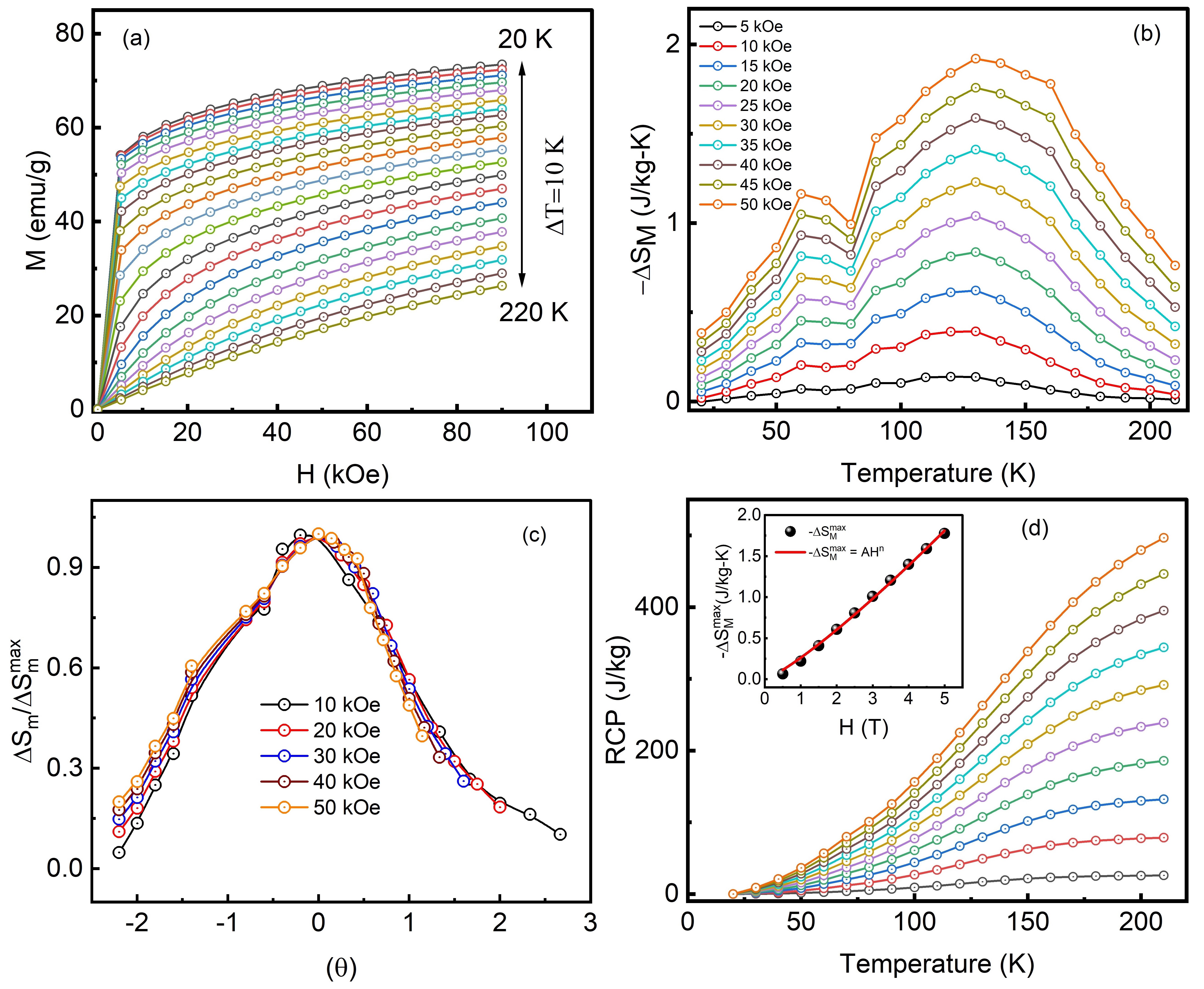}
    \caption{\label{fig:MCE}(a) The magnetic field dependent magnetization isotherms in the range 20 K - 220 K with $\Delta$T = 10 K. (b) Change in the magnetic entropy as a function of temperature for different magnetic fields. (c) Universal nature of normalized change in the magnetic entropy as a function of rescaled temperature ($\theta$) at different magnetic fields. (d) The temperature dependent RCP at different fields. Inset represents the maximum change in the magnetic entropy as a function of magnetic field along with the fit to power law equation and the exponent n = 1.19 indicates the SO-MPT of the alloy.}
\end{figure*}

Al$_2$MnFe shows a PM - FM phase transition without any noticeable hysteresis between the ZFC and FC curves around T$_C$, which might be useful for magnetic refrigeration (MR) applications. To have a deeper understanding about the order of MPT and MCE, MH isotherms in the temperature range of 20 K to 220 K ($\Delta$T = 10 K) are measured and presented in Fig. \ref{fig:MCE}\hyperref[fig:MCE]{(a)}.  Generally, MCE is described by two important parameters: (i) the isothermal magnetic entropy change ($\Delta$S$_M$), which occurs as a result of the change in ordering of magnetic moments with varying magnetic field, and (ii) the relative cooling power (RCP) associated with area under the magnetic entropy change curve (the amount of heat transferred during the magnetic refrigeration cycle). Using MH isotherms, the magnetic entropy change ($\Delta$S$_M$) may be calculated using Maxwell's relation given as \cite{ref53}:
\begin{align}
    \Delta S_M (T,\Delta H)&=S_M (T,H_F )-S_M (T,H_I )\nonumber\\
    &=\int_{H_I}^{H_F}(\partial M(T,H)/\partial T)_H dH,
    \label{eqn:20}
\end{align}
where H$_I$ and H$_F$ are the applied magnetic field with H$_I$ $<$ H$_F$ and $\Delta$H = H$_F$ - H$_I$. To determine $\Delta$S$_M$, the MH isotherms data are numerically integrated across discrete magnetic fields for different temperature intervals using the following equation:
\begin{equation}
    \Delta S_M (T,H)=\sum_i\frac{M_{i+1}(T_{i+1},H_i)-M_i(T_i,H_i )}{(T_{i+1}-T_i )}\Delta H_i,
    \label{eqn:21}
\end{equation}
where the experimentally recorded magnetization at T$_i$ and T$_{i+1}$ under magnetic field H$_i$ are represented by M$_i$ and M$_{i+1}$, respectively. $\Delta$S$_M$ as a function of temperature, extracted at various magnetic fields, based on equation (\ref{eqn:20}) is shown in Fig. \ref{fig:MCE}\hyperref[fig:MCE]{(b)}. The maximum value of $\Delta$S$_M$ increases with increasing magnetic field and a value of -$\Delta$S$_M^{max} \sim$ 1.92 J/kg-K at 50 kOe is observed. The $\Delta$S$_M^{max}$ appears in the vicinity of the transition temperature, as expected for a magnetic material. This is attributed to the fact that the maximum magnetic entropy change occurs at the ordering temperature of a typical FM. The nature of MPT can be determined by the analysis of normalised entropy change \cite{ref54}. In this approach, the entropy change corresponding to different magnetic fields is normalised with reference to the maximum entropy change and presented as a function of rescaled temperature ($\theta$). The temperature is rescaled according to the following relation \cite{ref53, ref55}:
\begin{equation}
    \theta=
    \begin{dcases}
        -\frac{(T-T_{max})}{(T_{r1}-T_{max})};&T<T_C\\
        \frac{(T-T_{max})}{(T_{r2}-T_{max})};&T>T_C
    \end{dcases}
    ,
    \label{eqn:22}
\end{equation}
where T$_{max}$ is the temperature corresponding to maximum value of  $\Delta$S$_M$, and T$_{r1}$ and T$_{r2}$ represent the reference temperatures below and above T$_C$, respectively. Fig. \ref{fig:MCE}\hyperref[fig:MCE]{(c)} illustrates the normalised change of $\Delta$S$_M$/$\Delta$S$_M^{max}$ as a function of rescaled temperature. It is evident that all the normalised entropy-change curves collapse into a single universal curve, thus confirming the SO-MPT in the studied alloy \cite{ref53}. Further, to check the field dependency of $\Delta$S$_M^{max}$, a power law given by $\Delta$S$_M^{max}$ = aH$^n$ (where a is a constant and n is an exponent) is employed to ascertain the order of MPT (shown in the inset of Fig. \ref{fig:MCE}\hyperref[fig:MCE]{(d)}). A value of exponent n lower than 2 (greater than 2) signifies a second (first) order MPT \cite{ref56}. The obtained value of exponent  n = 1.19 again confirms the SO-MPT in the prepared alloy.

Further, RCP can be determined from the area under the entropy change curve as function of temperature by the following equation \cite{ref54}:
\begin{equation}
    RCP=\int_{T_1}^{T_2}\vert \Delta S_M (T)\vert_{\Delta H} dT,
    \label{eqn:23}
\end{equation}
where T$_1$ and T$_2$ are the lower and upper limits of refrigeration cycle, respectively. Fig. \ref{fig:MCE}\hyperref[fig:MCE]{(d)} displays the RCP as a function of temperature. The maximum value of RCP at 50 kOe is 496 J/kg. The values of -$\Delta$S$_M^{max}$ and RCP together determine the quality of a magnetocaloric material and a higher RCP value is preferred for a better magnetocaloric performance. The obtained values of -$\Delta$S$_M^{max}$ and RCP for Al$_2$MnFe alloy are comparable to those of other Mn-Fe-Al based magnetocaloric materials as shown in Table \ref{tab:MCE} and obviously lesser than those of pure Gd \cite{ref57} and other rare-earth based magnetocaloric materials \cite{ref58}. However, it is imperative to state that the studied material shows significantly good magnetocaloric behavior among the materials comprising of abundantly available 3d transition metal elements with SO-MPT.
\begin{table}[h!]
    \caption{\label{tab:MCE}The maximum change in magnetic entropy (-$\Delta$S$_M^{max}$) and relative cooling power of Al$_2$MnFe HA compared with other magnetocaloric materials.}
    \begin{ruledtabular}
        \begin{tabular}{cccc}
            Sample & -$\Delta$S$_M^{max}$ & RCP & Reference\\
            & (J kg$^{-1}$ K$^{-1}$) & (J kg$^{-1}$) &\\
            \hline
            Gd & 9.7 & 556 & \cite{ref57}\\
            Fe$_{60}$Mn$_{15}$Al$_{25}$ & 0.95 & 395 & \cite{ref59}\\
            Fe$_{57.5}$Mn$_{17.5}$Al$_{25}$ & 1.07 & 430 & \cite{ref59}\\
            Mn$_{30}$Fe$_{20}$Al$_{50}$ & 2.6 & 300 & \cite{ref20}\\
            Mn$_{30}$Cu$_{20}$Al$_{50}$ & 3.1 & 270 & \cite{ref20}\\
            Al$_2$MnFe & 1.92 & 496 & This work\\
        \end{tabular}
    \end{ruledtabular}
\end{table}

\section{\label{sec:Conclusions}Conclusions}

In this report an extensive investigation of the reverse Heusler formula Z$_2$XY-type arc-melted polycrystalline Al$_2$MnFe is presented through DC magnetization and ACS measurements. The PXRD pattern for the alloy confirms its bcc-type B2 single phase nature having Pm$\Bar{3}$m space group. DC magnetization study suggests a second order PM - FM phase transition around T$_C$ = 122.9 K and another magnetic phase transition occurring at low temperature (around 14 K). The field dependent magnetization indicates a soft FM nature of the alloy and a strong itinerant type of FM exchange interactions as confirmed by RWR and SCR theory. The re-entrant cluster SG transition at low temperature region is confirmed by the frequency dependent shifting of peak temperatures from the ACS data analyzed using the critical slowing down model and Vogel-Fulcher law. The detailed analysis of the critical phenomenon around T$_C$ in the Al$_2$MnFe alloy estimates the critical exponents $\beta$ = 0.363, $\gamma$ = 1.384 and $\delta$ = 4.81 from MAPs, and $\beta$ = 0.37, $\gamma$ = 1.40 and $\delta$ = 4.78 from KFPs. The critical isotherm at T$_C$ yields the critical exponent $\delta$ = 4.75. The obtained critical exponents perfectly obey the scaling hypothesis and the isotherms below and above T$_C$ fall into two distinct branches of the universal curve, thus validating the consistency of the critical exponents. The critical analyses of the studied alloy thus suggest that the system is dominated by isotropic short-range exchange interactions, and the system belongs to the 3D-Heisenberg universality class, which is also confirmed from the range of interactions interpreted on the basis of renormalization group theory. The magnetocaoric effect of the alloy is studied in terms of the maximum magnetic entropy change and RCP whose values are 1.92 J/kg-K and 496 J/kg in a 5 T magnetic field, respectively. These values suggest that the alloy shows incredible magnetocaloric properties and poses tremendous potential for practical applications. The present work is expected to motivate the exploration of other reverse stoichiometric FHAs for unfurling their extraordinary physical properties that can be harnessed for advanced technological applications.

\begin{acknowledgments}
AKP acknowledges the support of SERB-DST, New Delhi, India (Grant no. CRG/2020/003590) and DST (Grant no. INT/RUS/RFBR/379). AKK acknowledges UGC New Delhi, India, for providing the financial support through a JRF Fellowship (16-6(DEC. 2018)/2019 NET/CSIR). Authors acknowledge the “Low temperature and high magnetic field facility" under CIF at the Central University of Rajasthan for magnetic measurements.
\end{acknowledgments}

\bibliography{references}

\end{document}

\title{Magnetic critical phenomena and low temperature re-entrant spin-glass features of Al$_2$MnFe Heusler alloy - Supplementary material}

\author{Abhinav Kumar Khorwal}
\affiliation{Department of Physics, Central University of Rajasthan, Ajmer - 305817, Rajasthan, India}

\author{Sujoy Saha}
\affiliation{Department of Physics, Central University of Rajasthan, Ajmer - 305817, Rajasthan, India}

\author{Mukesh Verma}
\affiliation{Department of Physics, Central University of Rajasthan, Ajmer - 305817, Rajasthan, India}

\author{Lalita Saini}
\affiliation{Department of Physics, Indian Institute of Technology Gandhinagar, Palaj, Gandhinagar - 382055, Gujarat, India}

\author{Suvigya Kaushik}
\affiliation{Department of Physics, Indian Institute of Technology Gandhinagar, Palaj, Gandhinagar - 382055, Gujarat, India}

\author{Yugandhar Bitla}
\affiliation{Department of Physics, Central University of Rajasthan, Ajmer - 305817, Rajasthan, India}

\author{Alexey V. Lukoyanov}
\affiliation{M.N. Mikheev Institute of Metal Physics UrB RAS, 620108, Ekaterinburg, Russia}
\affiliation{Institute of Physics and Technology, Ural Federal University, 620002, Ekaterinburg, Russia}

\author{Ajit K. Patra}
\email[Corresponding author: ]{a.patra@curaj.ac.in}
\affiliation{Department of Physics, Central University of Rajasthan, Ajmer - 305817, Rajasthan, India}

\maketitle

\renewcommand{\thefigure}{S\arabic{figure}}

\begin{figure}
    \includegraphics[width=0.95\textwidth]{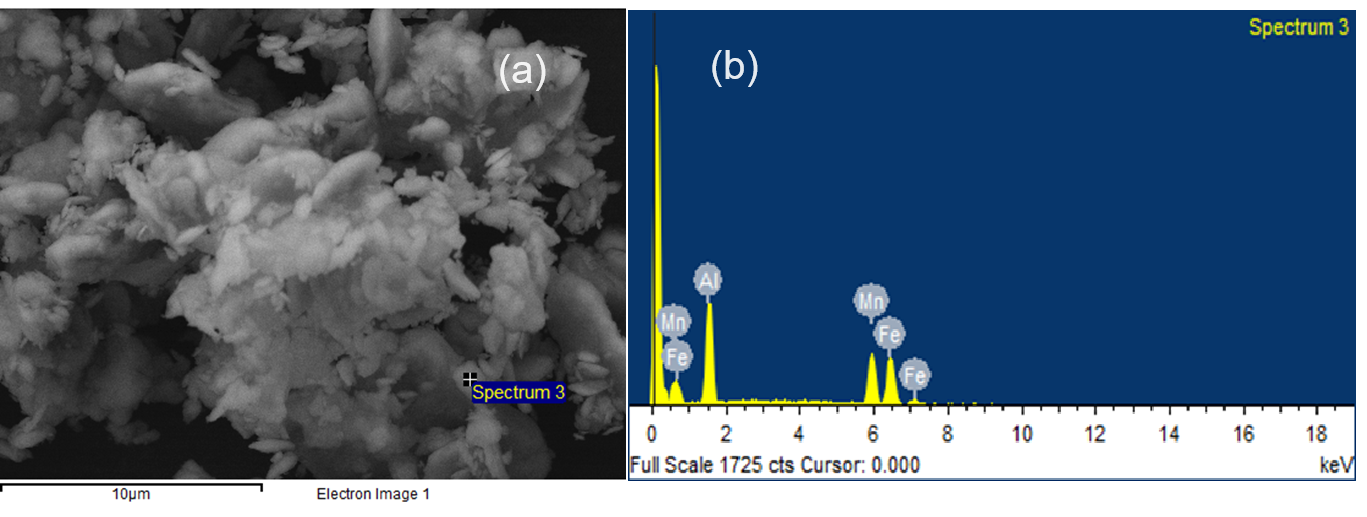}
    \caption{\label{fig:EDX}(a) SEM image, and (b) EDX spectra of prepared Al$_2$MnFe alloy to check the homogeneity and elemental composition.}
\end{figure}

\begin{figure}
    \includegraphics[width=0.75\textwidth]{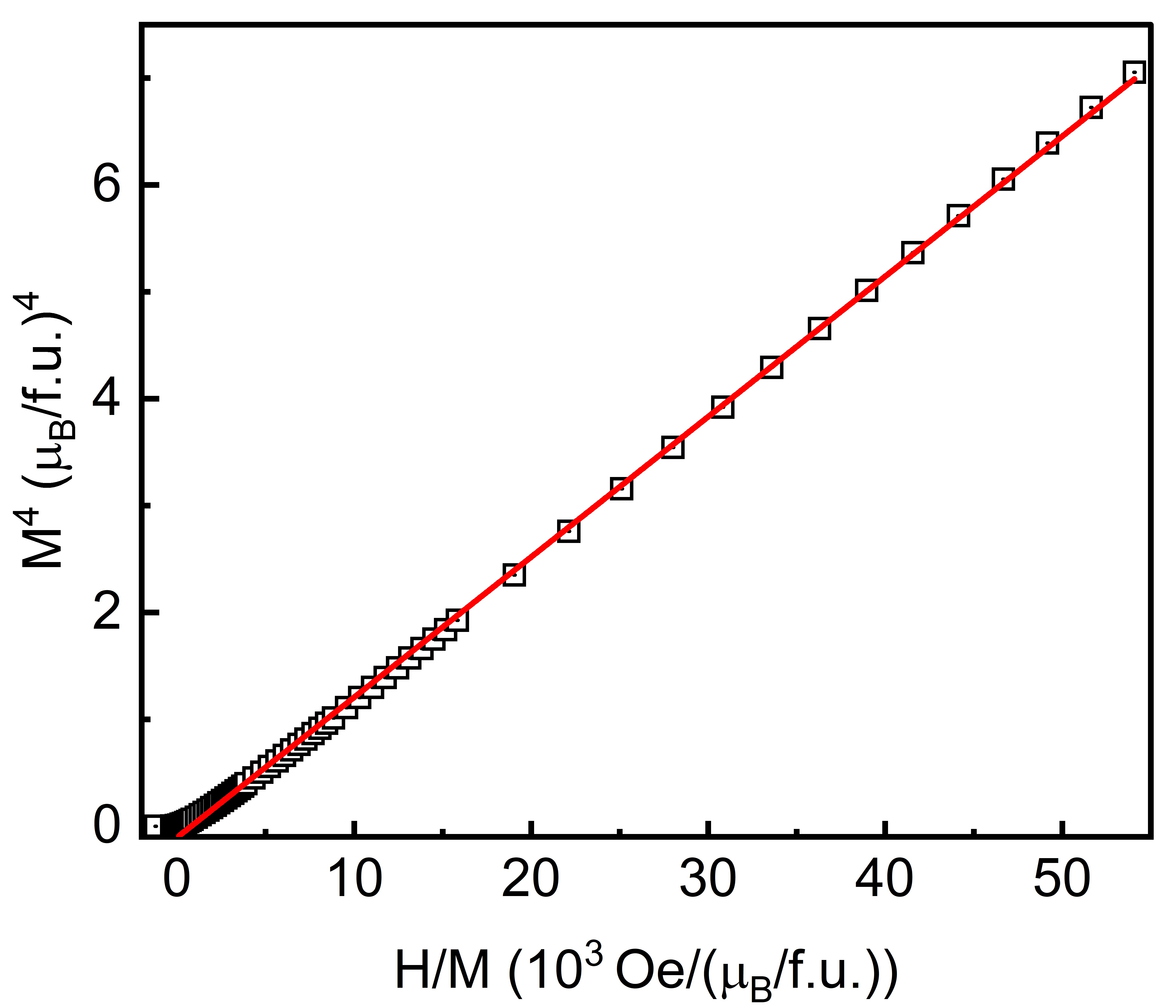}
    \caption{\label{fig:RW} The M$^4$ vs (H/M plot of Al$_2$MnFe Heusler alloy at T$_C$ along with the linear fit.}
\end{figure}

\begin{figure}
    \includegraphics[width=0.95\textwidth]{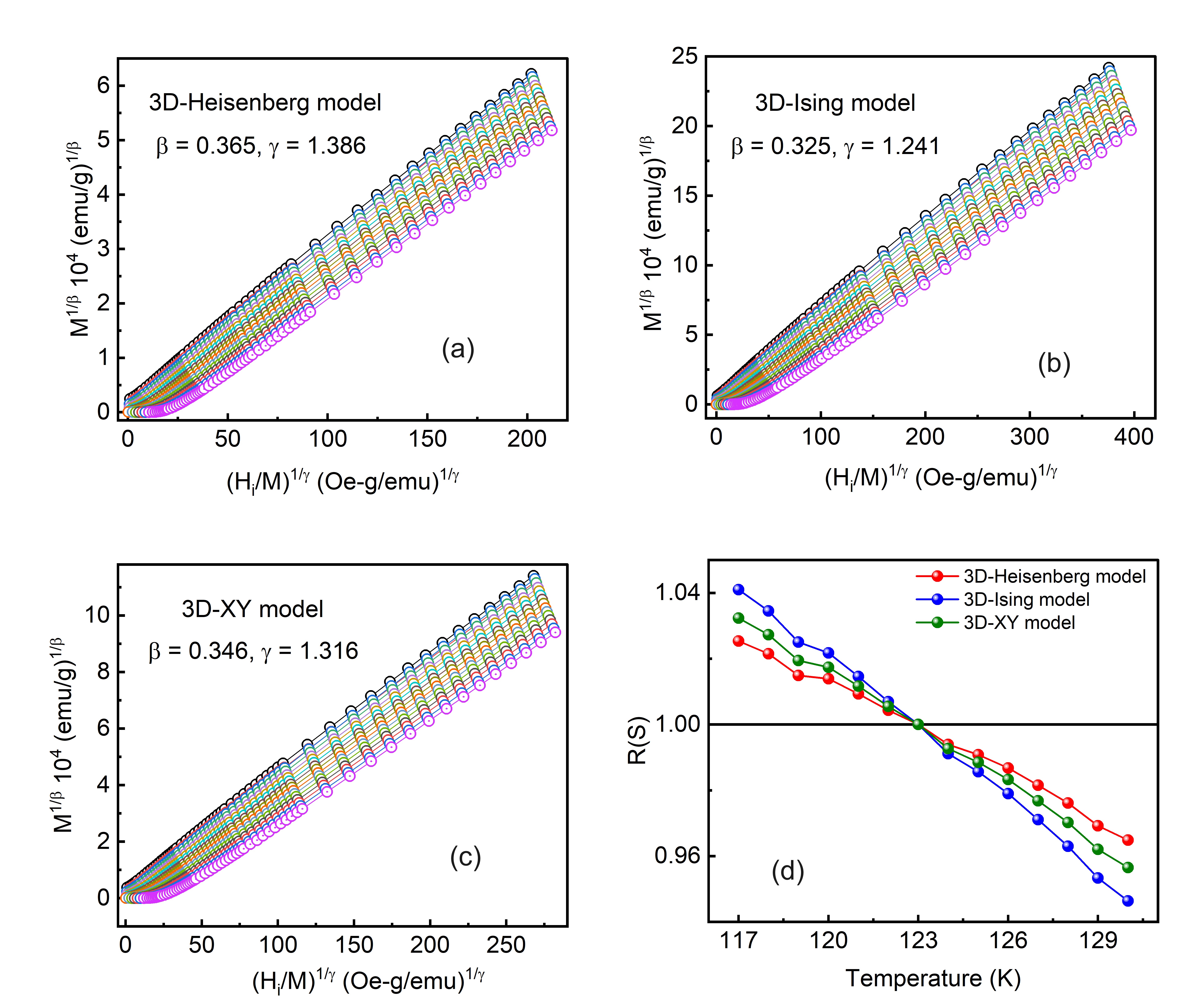}
    \caption{\label{fig:MAP}(a-c) Representation of Modified Arrott plots using various theoretical critical exponents predicted for three dimensional (3D) systems such as 3D-XY ($\beta$ = 0.346, $\gamma$ = 1.316), 3D-Heisenberg ($\beta$ = 0.365, $\gamma$ = 1.386) and 3D-Ising ($\beta$ = 0.325, $\gamma$ = 1.241) models. (d) The relative slope RS = S(T)⁄S(T$_C$) is plotted as a function of temperature for each model . For an ideal model, the value of RS is close to 1. It is observed that the 3D Heisenberg model has a relatively smaller deviation compared to the other models.}
\end{figure}